\newcommand{\bf}[1]{\textbf{#1}}
\documentclass[sigconf,screen,authorversion, nonacm]{acmart}

\AtBeginDocument{%
  }

\usepackage{subcaption}
\usepackage{float}
\usepackage{placeins}
\usepackage[markup=default,commandnameprefix=always, commentmarkup=footnote]{changes}
\definechangesauthor[name={AK}, color=orange]{AK}

\copyrightyear{2026}
\acmYear{2026}
\setcctype{by}
\acmConference[ICCAD '26]{IEEE/ACM International Conference on Computer-Aided Design}{November 08--12, 2026}{San Jose, CA, USA}
\acmBooktitle{IEEE/ACM International Conference on Computer-Aided Design (ICCAD '26), November 08--12, 2026, San Jose, CA, USA}
\acmDOI{10.1145/3831252.3834057}
\acmISBN{979-8-4007-2873-0/2026/11}
\begin{document}

\title{Routing Dense Layouts with History-Aware Offline Reinforcement Learning using LSTM}

\author{Afsara Khan}
\email{atk331@nyu.edu}
\affiliation{%
  \institution{New York University}
  \city{Brooklyn}
  \state{New York}
  \country{USA}
}

\author{Austin Rovinski}
\email{rovinski@nyu.edu}
\affiliation{%
  \institution{New York University}
  \city{Brooklyn}
  \state{New York}
  \country{USA}
}









\begin{abstract}
Detailed routing remains a dominant runtime bottleneck in physical design due to increasing complexity of design rules. Modern routers can struggle to resolve persistent violations under dense operating conditions.
While recent work leverages reinforcement learning (RL) to dynamically select costs for each routing iteration, we find that this technique struggles with high-density designs where routing solutions are significantly harder.
To address this, we present a history-aware offline RL policy which predicts iterative cost weights in these dense regimes to improve convergence across placement densities by utilizing readily available features from the router.
Our policy uses conservative Q-learning similarly to prior work; however, our key insight is that \textit{addition of a lightweight LSTM architecture and additional features can retain sequence context and improve routing convergence across multiple densities and route guide qualities.}
Our policy can be integrated into any cost-based router with minimal pipeline changes, as it does not interfere with the core search algorithm. 
We evaluate our policy on held-out density and adjustment settings, including difficult operating points induced by dense placement and low guide quality. Our policy reduces design rule violations (DRVs) by an average of 92\% over the top public baseline while simultaneously reducing runtime by 10\%.
\end{abstract}




\maketitle

\section{Introduction}

Detailed routing is often a very time-consuming step in the place \& route flow. As technology nodes become more costly, achieving high logic density is critical to both reducing cost and improving performance. Modern routers rely on cost-driven iterative search and repair pathfinding algorithms, where the relative weighting of cost penalties can strongly influence the convergence behavior.
Most public state-of-the-art work relies on fixed or manually designed cost schedules~\cite{chen2020detailed,kahng2018tritonroute,kahng2020tritonroute}, where the algorithm continuously rips up and reroutes until it finds a set of costs that route cleanly, or it terminates with violations. 

As placement density increases, so too does the routing congestion. This leaves fewer open tracks available for the detailed router to perform rerouting, which increases the likelihood of stalls and persistent violations. Fixed cost schedules and simple regressions struggle with high density placements, because the best routing solution varies dramatically based on the local congestion and the current mix of violations, both of which can change from one routing iteration to another. While tuning static costs can help improve results on a given design family or operating point, the cost map can be brittle and make other operating points worse. 

In this paper, we study iteration-level control of existing routing cost weights in high utilization regimes, when the detailed router is left with very limited resources for rerouting. We demonstrate our implementation using OpenROAD as a public reference point; however, any iterative cost-based detailed router can adopt the same technique with router-specific feature extraction and lightweight software integration. The broader methodology of using a learned policy to control routing cost weights at each iteration using built-in router features can be extended to dense regimes where open-source state-of-the-art routers fail, directly resolving layouts that are difficult to route with existing routing flows.

\begin{figure}
    \centering
    \includegraphics[width=1\linewidth]{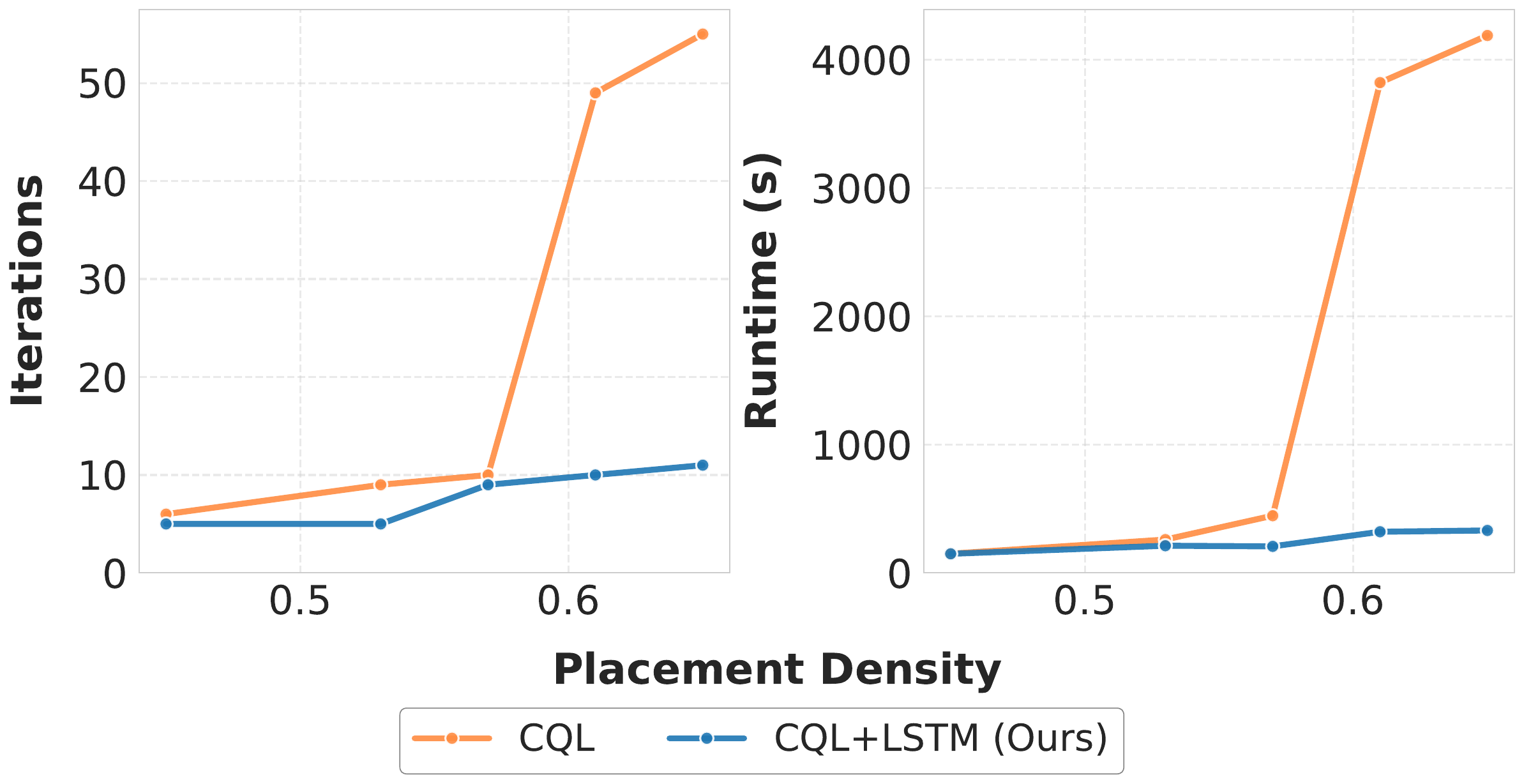}
    \caption{Routing performance for \texttt{aes} vs. placement density. CQL denotes the approach from prior work \cite{khan2026accelerating} and CQL+LSTM is this work.} \vspace{-0mm}
    \label{fig:model_comparison}\vspace{-0mm}
\end{figure}

Prior work~\cite{khan2026accelerating} has shown that offline conservative Q-learning (CQL) can act as an intelligent cost weight generator and accelerate convergence in a fixed operating regime with superior results over public baselines. However, varying density and operating conditions introduce additional challenges, since layout features can change substantially even with small density shifts, causing distribution shift between training and inference and pushing cost weight actions toward saturation. Alongside introducing an LSTM head for stronger sequential awareness, our method addresses these distributional issues through feature transformation, masking of density and adjustment inputs during training, and the use of ratios and logarithms for local signals whose absolute counts can vary widely. We demonstrate in Figure~\ref{fig:model_comparison} that, while the CQL approach of prior work~\cite{khan2026accelerating} degrades sharply as placement density rises, our model remains effective in these higher utilization regimes.

The results show consistent improvements over a wide range of densities. The largest gains appear when a test density lies between training points, which is consistent with the broader tendency for learned models to behave more reliably in interpolation than in extrapolation~\cite{dakhmouche2025neuralnetworksmasterextrapolation}.
However, the interpolated densities remain nontrivial in our setting, since even intermediate density values can induce materially different layouts and routing behavior (discussed further in Section~\ref{sec:ExperimentalResults}).

We train a conservative Q objective with a stochastic actor and add an LSTM head for the policy to better retain short routing history to link recent choices to their effects rather than react to a single snapshot. Each run is subject to strict limits: at most 65 routing iterations or 7 days of runtime. Runs that exceed either limit are not considered and we avoid extrapolating outcomes. Our work offers the following contributions:

\begin{itemize}
\item  Development of a detailed routing training dataset using perturbation sampling and random exploration techniques across various density and adjustment configurations, with emphasis on dense routing regimes
\item  Identification of critical routing state variables and reward components suitable for history-aware iterative control in detailed routing using reinforcement learning (RL)
\item  Implementation and evaluation of an offline RL approach using conservative Q-learning with LSTM based soft actor-critic network to predict improved cost weights for detailed routing that outperform public baselines.
\item  Demonstration of improved routing convergence for diverse designs in high utilization regimes.
\item The training and inference implementation are open sourced at \url{https://github.com/realise-lab/RLDRT}.
\end{itemize}

The remainder of the paper is organized as follows.
Section~\ref{sec:RelatedWork} discusses related work,
Section~\ref{sec:Methodology} discusses our methodology, Section~\ref{sec:ModelArchitecture} discusses our model architecture and training, Section~\ref{sec:ExperimentalResults} discusses the experimental results, and Section~\ref{sec:Conclusion} provides our conclusions.

\section{Related Work}
\label{sec:RelatedWork}

\subsection{Detailed Routers}
The foundation of modern detailed routing traces back to Lee's maze routing algorithm \cite{lee1961algorithm}, a breadth-first search (BFS) that guaranteed minimum cost paths. Rapid notable improvements on this base algorithm included the A* search using heuristics to guide the search \cite{arnold,bidir}, along with techniques like line search~\cite{hightower1969solution} to speed up execution. Core strategies in modern routers include iterative rip-up and reroute \cite{kahng2020tritonroute}
and sometimes leverage multicommodity flow concepts \cite{han2015}. TritonRoute \cite{kahng2020tritonroute} utilizes the prior findings to employ an iterative A*-based search with partitions, enhanced by dynamic boundary adjustments. Similarly, Dr. CU~\cite{chen2020detailed} uses Dijkstra's algorithm coupled with a sparse grid-graph and partitioning to efficiently route nets. As newer technology nodes arrive, routing research has shifted focus towards techniques like gridless pin access \cite{nieberg2011}, routing under complex patterning like SADP \cite{ding2017,liu2014}, minimum area-sensitive path search \cite{ahrens2015detailed}, and ILP-based formulations \cite{han2015}.

In terms of academic routers, OpenROAD~\cite{ajayi2019openroad-gomactech} provides state-of-the-art performance, achieving 0 DRVs on the ISPD `18~\cite{mantik2018ispd} benchmark and 0 DRVs on all but one test case on ISPD `19~\cite{liu2019ispd}. OpenROAD's router is derived from TritonRoute~\cite{kahng2018tritonroute,kahng2020tritonroute}, but several enhancements have been made and it is actively maintained to continue improving detailed routing results.

In the academic routing flows discussed above, the exposed cost weights that guide the iterative search are controlled by manually designed schedules rather than by the evolving routing state of the design.

\subsection{Machine Learning Techniques in Routing}


Several prior works have shown the benefits of applying traditional machine learning to global and detailed routing. Many works have focused on congestion and DRV hotspot prediction at the routing level~\cite{10,11}, while others have focused on predicting and avoiding pin access violations~\cite{12}. These predictions typically required secondary mechanisms to influence routing behavior. More direct guidance has been explored via reinforcement learning. For example, Chen et al. \cite{13} propose an online RL framework that combines graph neural networks (GNNs) with PPO policy learning.

Prior work \cite{khan2026accelerating} demonstrates that offline RL can learn effective cost weights to accelerate convergence in detailed routing. However, that work trains and evaluates at a single operating point. Density and adjustment changes shift the underlying feature distribution substantially, and a policy fit to a single density can no longer sustain convergence under that shift without deliberate feature and architectural engineering. Retaining the offline RL formulation from prior work~\cite{khan2026accelerating}, we introduce robust feature construction and transformation, masking of density and adjustment signals during training, and an LSTM that conditions on recent routing history rather than a single state. These additions fundamentally change the nature of the problem: from accelerating convergence at one operating point to sustaining it across dense layouts, including hard cases that do not converge at all under existing flows.

While we demonstrate our formulation using OpenROAD, the technique can be adapted to any cost based router.
To the best of our knowledge, no prior work has studied history aware offline RL for iterative cost weights across varying routing regimes in general detailed routing.

\section{Methodology}
\label{sec:Methodology}

\subsection{Routing and Policy}

We study detailed routing across varying placement utilization, with special attention to the difficult, higher-density cases. Global routing adjustment controls how aggressively the global router is allowed to use the available routing tracks. At 0.0, the global router may use all tracks, which packs nets tightly and leaves the detailed router with very limited resources to resolve violations. Higher adjustment restricts a higher fraction of track capacity from the global router, leaving more usable space for the detailed router that follows \cite{kahng2020tritonroute}. Lowering the global-routing adjustment is not universally required as density increases; it becomes necessary only when, at a given density, the global router would otherwise return a congested guide. In those situations, reducing adjustment leaves the global router with more usable tracks to resolve congestion and return a routing guide for the detailed router. However, this leaves fewer reserved detailed routing resources, as mentioned above, increasing the likelihood of stalls and persistent violations.

The policy interfaces through the standard cost multipliers. At the end of iteration \(k\), the router aggregates a global feature snapshot to calculate the feature vectors for the model.
The agent then returns a shared set of cost multipliers for iteration \(k{+}1\): \textit{drcCost}, \textit{markerCost}, \textit{fixedShapeCost}, \textit{markerDecay}. These multipliers control TritonRoute's routing search by acting as penalty weights: \textit{drcCost} and \textit{markerCost} penalize routing through Design Rule Check (DRC) violating regions and near existing violation markers, \textit{fixedShapeCost} penalizes overlap with fixed obstructions, and \textit{markerDecay} sets how quickly old violation markers lose influence \cite{kahng2020tritonroute,khan2026accelerating}. Rather than fixing these multipliers to a schedule as seen in the baseline router, our policy resets them each iteration from the current routing state. We adopt global iteration control to keep integration lightweight and avoid additional coordination and overhead that per-partition control would require. At iteration \(0\), the router uses the default cost weights because no prior state history is yet available. The core search remains unchanged. 

Iteration count and total runtime are related but not identical. Each iteration activates multiple partitions across workers. When violations are stubborn, each partition requires longer time to rip up and repair its local region, so the iteration takes longer; easier violations finish faster.
We tested cumulative DRV as a runtime proxy by correlating violation traces with measured time across designs and operating points. We define the cumulative count in Equation~\ref{eq:cumulative} as
\begin{equation}\label{eq:cumulative}
S_k \;=\; \sum_{t=0}^{k} \mathrm{DRV}_t,
\end{equation}
where \(\mathrm{DRV}_t\) is the violation count at iteration \(t\). The correlation was inconsistent: per-iteration DRV often showed only weak association with per-iteration runtime, and runs with similar total DRV sometimes had large runtime differences because a small number of late-stage, stubborn violations dominate cost. In addition, the router resolves multiple partitions in parallel \cite{kahng2020tritonroute}, so a larger number of violations does not necessarily increase runtime linearly.

\subsection{Data Generation}

As with most reinforcement learning methods, data generation is central to performance. We generate data from 8 Nangate45 designs in the OpenROAD Design Suite~\cite{rovinski2020bridging} and span 6 placement densities per design. The densities are evenly spaced and start at the default utilization for the design suite. Global-routing adjustment varies from 0.0 to 0.3. Lower adjustments yield valuable but slow traces. Higher adjustments yield many more traces and can run in parallel with others. Most data therefore come from 0.10 to 0.30, with a smaller portion at 0.00 where runs are slower but informative. In total, about 10,000 routing runs were collected across designs and densities. Harder routing cases in the dataset arise naturally from plausible density and adjustment sweeps within the same routing flow, where higher density increases routing pressure and lower adjustment is sometimes needed to obtain a routable guide. 

\begin{figure}[htbp]
    \centering
    
    \begin{subfigure}{0.48\columnwidth}
        \includegraphics[width=\linewidth]{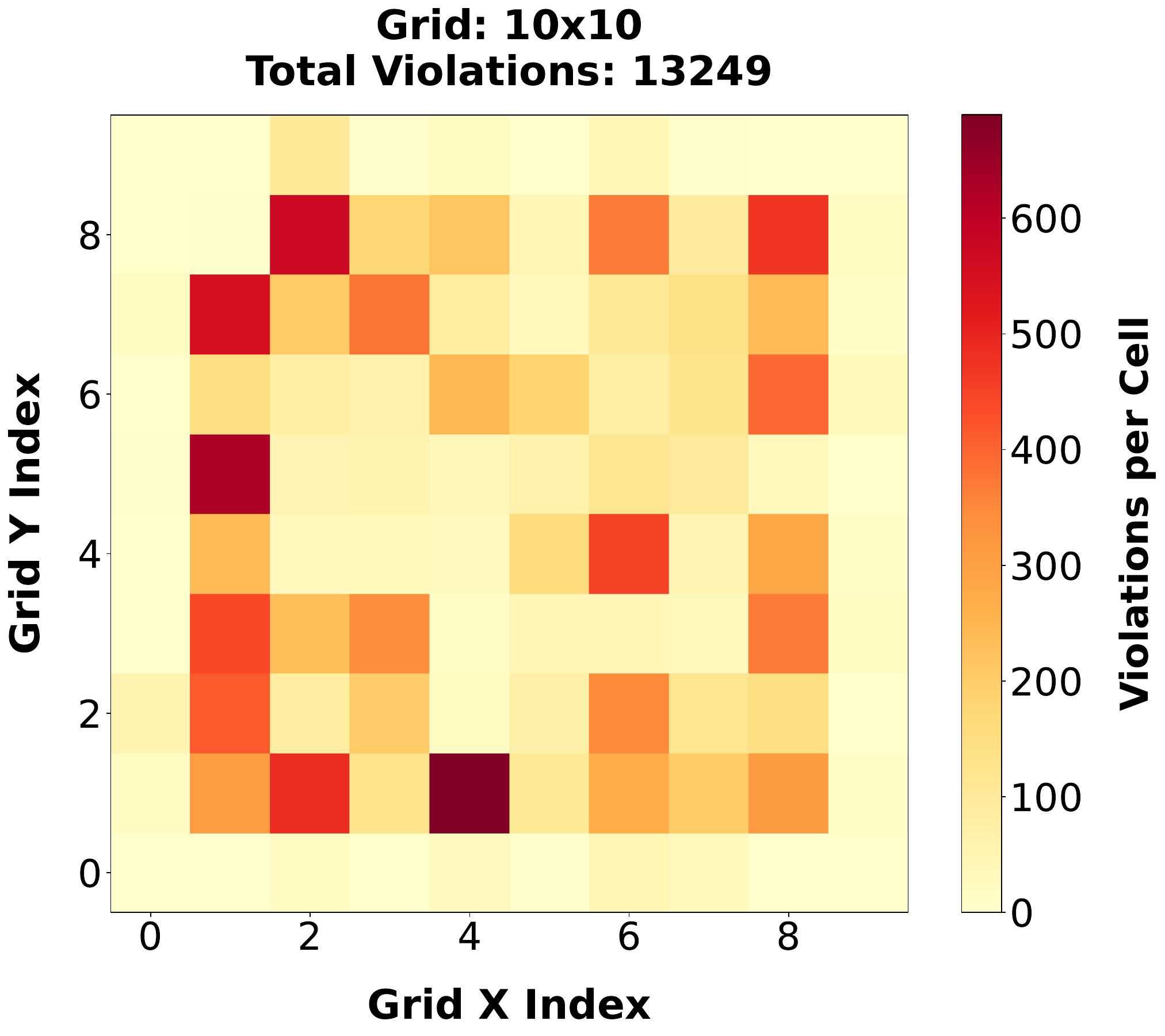}
        \caption{Iteration 0}
    \end{subfigure}%
    \hfill
    \begin{subfigure}{0.48\columnwidth}
        \includegraphics[width=\linewidth]{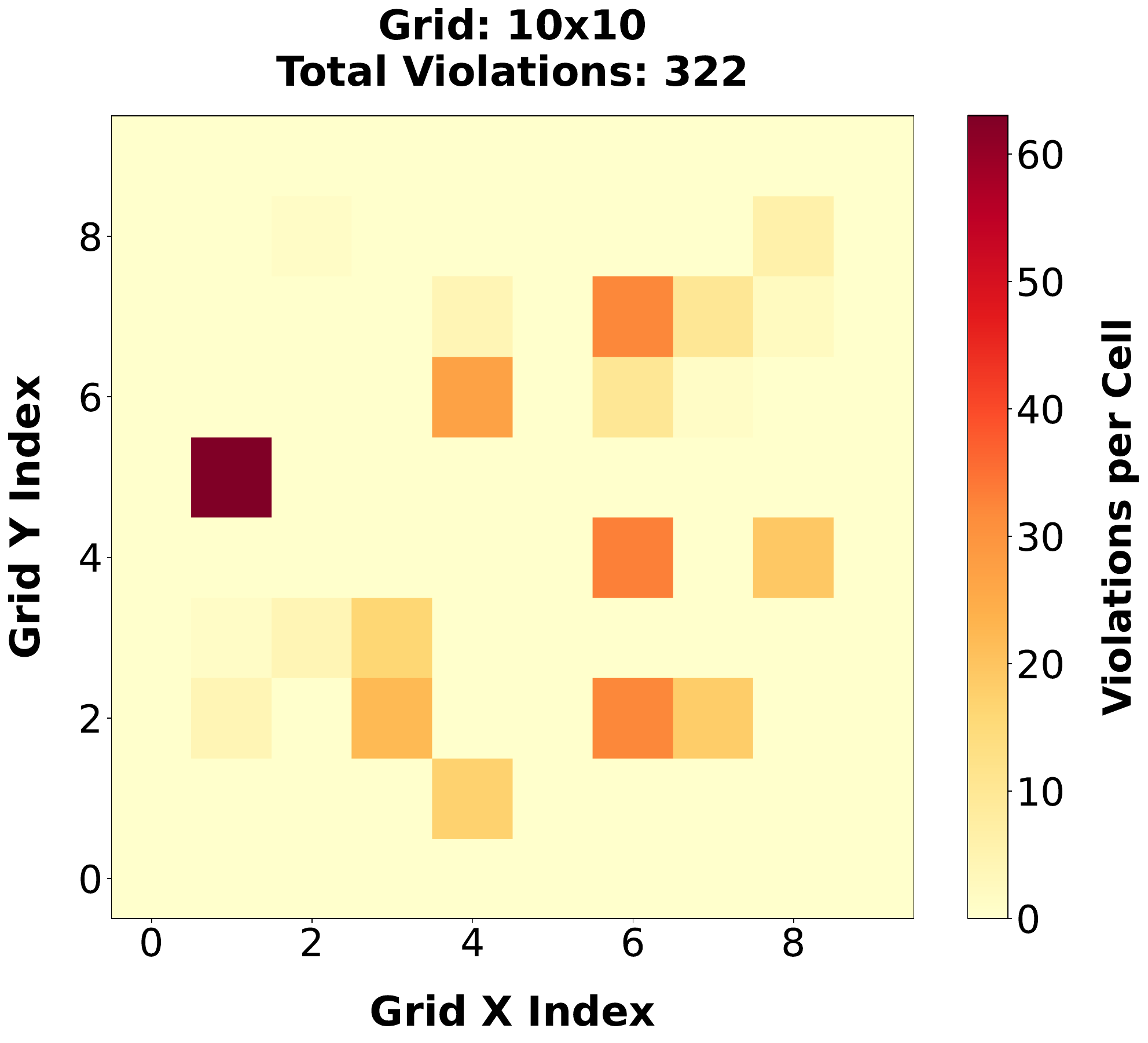}
        \caption{Iteration 4}
    \end{subfigure}
    \caption{Example violation progress on \texttt{aes} in an 11-iteration routing run}
    \label{fig:2}
\end{figure}

A coarse grid over the die records a violation heatmap at each iteration. The grid also captures simple dynamics such as whether violations remain in place, migrate, or form clusters as routing proceeds. Figure~\ref{fig:2} shows one region at iteration 0 and iteration 4. All signals come from the detailed router or from light transforms of those signals, so feature extraction overhead remains minimal relative to routing runtime, especially on complex designs.

Feature values are normalized to a common scale based on the training statistics. For features whose raw magnitudes can vary substantially across different routing settings, we prefer ratios or logarithmic transforms over raw counts. For example, current and initial DRV are log transformed, and progress is represented relative to the initial DRV from the first iteration. This practice provides a more stable signal across various densities and prevents the model from action saturation due to drastic distributional shifts in data.
The action itself is part of the state: we include the previous iteration weights and their change rate. We categorize features into two types:
\textbf{Dynamic Features}:
{\begin{itemize}
    \item Current violation count (normalized)
    \item Initial violation count (normalized)
    \item Stagnant violation regions which are based on coarse-grid cells whose aggregate local violation count does not decrease across three consecutive iterations
    \item Change in the maximum local violation count and clustering spread for the current iteration, where clustering spread measures how unevenly violations are distributed across active coarse-grid cells
    \item Violation type ratios for common classes: short, metal spacing, cut spacing, end-of-line spacing
    \item Previous weights, i.e. drcCost, markerCost, fixedShapeCost, markerDecay 
\end{itemize} }

\clearpage 
\textbf{Static Design Features}:
\begin{itemize}
    \item Terminal count in log scale
    \item Die area in log scale
    \item Placement density
    \item Routing adjustment
    \item Interaction of density, adjustment, and terminal count
 (the goal is to signal that the same density can behave differently by design, and that the hardest cases arise from combinations such as high density with 0.0 adjustment and high terminal count) 
    \end{itemize}

The underlying router parameters follow prior work~\cite{kahng2020tritonroute}, while the derived coarse grid and progress features are contributions introduced in this work.

After each iteration, we sample each of these values as part of the state $s$. A ``sequence'' $S$ is formed by a series of states $\{s_1,s_2,...,s_n\}$, where $n$ is the number of iterations and forms from one full routing run that terminates either at DRV convergence or at the iteration/runtime cap. Each sequence forms a data point which is used to train our RL model.

\section{Model Architecture}
\label{sec:ModelArchitecture}

We frame weight selection as a sequential decision process. At the end of iteration \(k\), the routing engine emits a state vector \(s_k\) and the policy returns the cost multipliers for iteration \(k{+}1\). The episode starts at iteration \(0\) and terminates when DRV reaches \(0\) or when iteration exceeds \(65\).

\noindent\textbf{Formulation.}
We adopt offline reinforcement learning with a conservative Q objective as described by \cite{cql}. The network is based on a stochastic actor in the soft actor-critic (SAC) family with LSTM heads, and we model state as a sequence. Both policy and value functions consume short temporal windows so the agent can condition on recent routing history.  The window length is a tuned hyperparameter (Table~\ref{tunetable}). Within an episode, the hidden state propagates across iterations and resets only between episodes, so the effective context reaches beyond the window itself.

\noindent\textbf{State representation.}
The state contains the signals described in Section~\ref{sec:Methodology}, including the previous iteration’s weights, and is fed to sequence encoders before the multilayer heads. An episode starts at iteration \(0\) and ends when DRV reaches \(0\) or when iteration exceeds \(65\). This sequence-aware formulation is a key change relative to non-recurrent baselines.

\noindent\textbf{Stability of standard Q-learning.}
We also trained a standard Q-learning variant by setting the conservative weight to zero. Despite a broad hyperparameter search with Optuna over learning rates, discount, target update rate, batch size, entropy temperature, and gradient clipping, losses remained unstable. We observed sustained growth in the Bellman error, frequent gradient explosions, and pronounced overestimation on out-of-distribution actions drawn from the replay buffer. Double-Q critics with target networks slowed divergence but did not prevent it. The conservative Q objective provided the needed regularization and produced stable training curves on the same data.

\noindent\textbf{Policy and critics.}
Both actor and critics are sequence aware. Each network uses \(2\) LSTM layers followed by a multilayer head. Hidden state propagates across iterations within an episode and resets at the start of the next episode. This captures short-range dependencies such as violation migration, hotspot persistence, and delayed effects of weight changes that may appear one or two iterations later.

\noindent\textbf{Action space and safety.}
The action is a 4-dimensional vector that specifies the next iteration’s multipliers \{\texttt{drcCost}, \texttt{markerCost}, \texttt{fixedShapeCost}, \texttt{markerDecay}\}. Outputs pass through a bounded squashing function into action ranges matched to the training data. This mapping avoids pathological values and keeps the downstream search stable.

\noindent\textbf{Learning objective and Reward Function.}
We validated the chosen reward function offline on the collected dataset by applying to logged trajectories. The reward scalar clearly distinguished runs with the lowest iteration counts and the fastest average runtimes from the rest of the dataset. Motivated by the routing analysis in Section~\ref{sec:Methodology}, we prioritize DRV reduction with an iteration penalty and hotspot term based on changes in the maximum local violation count, since violation \emph{difficulty} rather than count often drives runtime. The reward aggregates the following: \begin{itemize}
  \item \textbf{Progress:} improvement rate relative to the initial DRV.
  \item \textbf{Speed:} a convergence bonus when $\mathrm{DRV}=0$ and an iteration penalty that grows with $k$, so earlier completion earns higher return.
  \item \textbf{Difficulty-aware scaling:} bonuses and penalties are scaled by a design or instance complexity factor built from log transformed terminal count and die area, together with a bounded density modulation, that may be expressed as: 
    \begin{equation}\label{eq:cinst}
    C_{\mathrm{inst}} = \left(1 + \alpha \log(T+1) + \beta \log(A+1)\right) g(\rho),
    \end{equation}
  
  where $T$ is terminal count, $A$ is die area, $rho$ is placement density, and $g(\rho)$ is a bounded modulation term. This scaling ensures that identical absolute gains receive larger credit on harder instances.
  \item \textbf{Locality:} an explicit hotspot term based on changes in the maximum local violation count on the coarse grid.
\end{itemize}

We intentionally \emph{do not} include cumulative DRV in the training objective, and we do not use true runtime during training, since data generation runs in parallel across machines with variable load. Final runtime claims are based on isolated validation runs as stated in Section~\ref{sec:Methodology}.

\noindent\textbf{Training protocol.}
The replay buffer stores complete episodes and contiguous sequences so temporal order is preserved; within-episode shuffling is disabled. 

We regularize density and adjustment interactions with feature masking (feature dropout) on the input layer so the policy does not overfit to a single operating point and learns to adjust to distribution shifts \cite{featuremask}. For selected features, we form a masked view
\begin{equation}\label{eq:mask}
\tilde{\mathbf{x}} \;=\; \mathbf{x} \odot \mathbf{m}, \quad m_i \sim \mathrm{Bernoulli}(1-p_i),
\end{equation}
where \(p_i \in [0,0.5]\) is the masking probability tuned during training and only a small subset of inputs is eligible for masking (placement density, adjustment signals, and their interaction). Masking is applied to training episodes only. Validation and test use full features. We search \(p_i\) with Optuna over \([0,0.5]\) per masked feature and select the configuration that yields the best isolated runtime and iteration metrics on the validation set.

Optimization uses Double-Q critics with target networks and a soft policy update. Inputs are normalized; ratios replace raw counts when ranges are wide; logs are used when growth is steep.


\begin{figure}
    \centering
    \label{RLflow}
    \begin{subfigure}{0.5\columnwidth}
        \includegraphics[width=\linewidth]{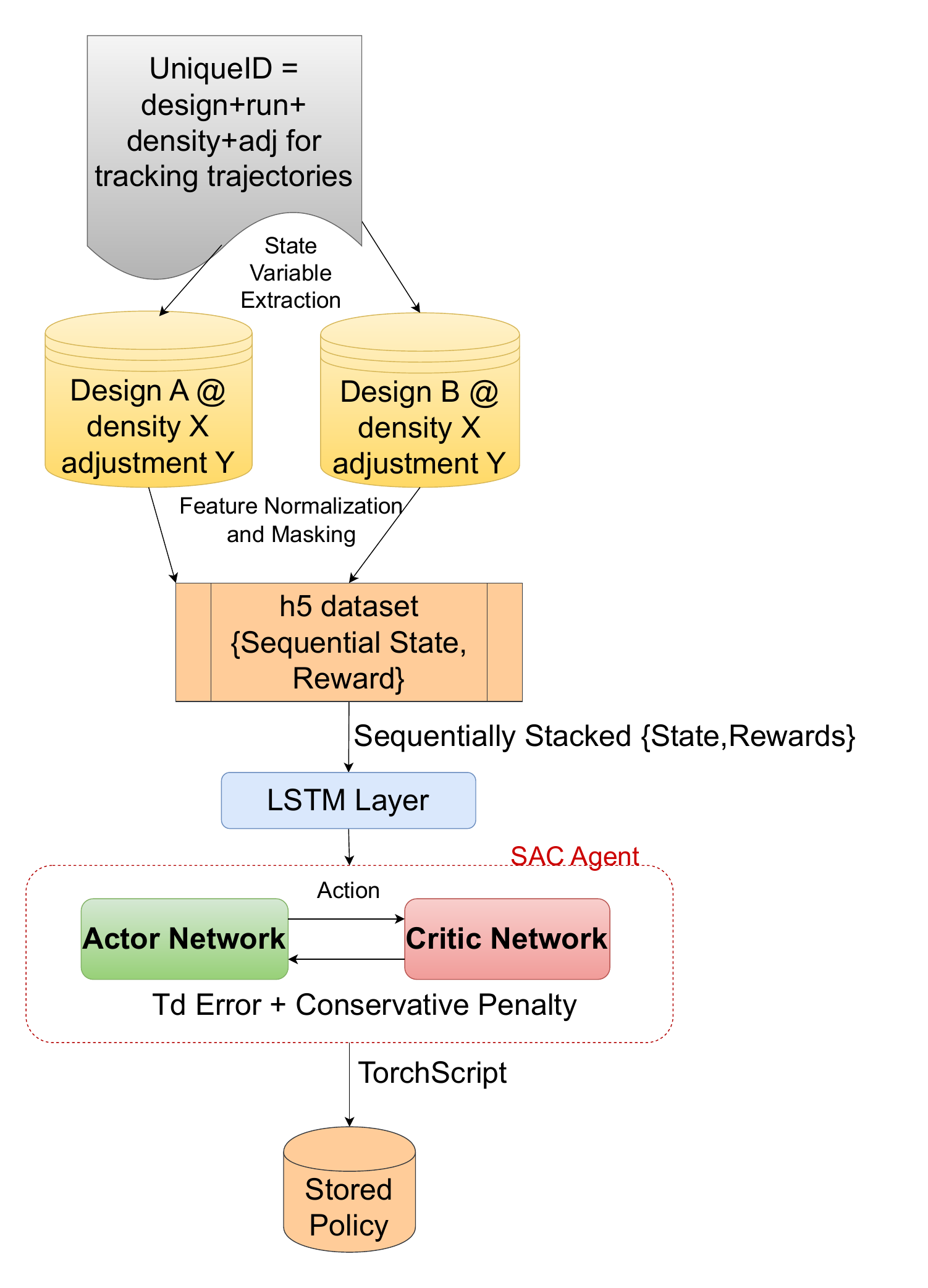}
        \caption{Training}
    \end{subfigure}%
    \hspace{0.5mm} 
    \begin{subfigure}{0.425\columnwidth}
        \includegraphics[width=\linewidth]{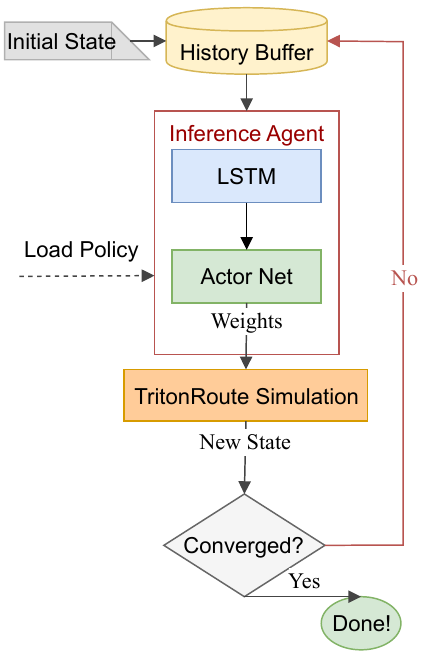}
        \caption{Inference}
    \end{subfigure}
    \caption{Training and Inference Summarized}
\end{figure}

\begin{table}
\begin{center}
\caption{CQL Hyperparameter Values}
\label{tunetable}
    \begin{tabular}{|c|c|} \hline 
        \textbf{Hyperparameter} &  \textbf{Value} \\ \hline
 \texttt{Critic and LSTM Dropout}&$1.00 \times 10^{-1}$\\
 Temporal Window (iters)&8\\\hline\hline
 \texttt{Critic Hidden Units}&512, 512, 256\\\hline 
        \texttt{Actor Learning Rate} & $1.00 \times 10^{-4}$\\ \hline 
        \texttt{Critic Learning Rate} & $3.00 \times 10^{-4}$\\ \hline 
        \texttt{Conservative Weight}& $2.44 \times 10^{0\phantom{-}}$\\ \hline 
        \texttt{Batch Size} & 512\\ \hline 
        \texttt{Initial Temperature} & $1.00 \times 10^{0\phantom{-}}$\\ \hline 
        \texttt{Temperature LR} & $1.00 \times 10^{-4}$\\ \hline 
        \texttt{Tau (Polyak $\tau$)} & $1.00 \times 10^{-2}$\\ \hline 
    \end{tabular}
    \vspace{-0mm}
\end{center}    
\end{table}

\noindent{\textbf{Model Parameters and Hyperparameters.}} Our goal in tuning was to accelerate convergence in favorable weight regimes while avoiding overestimation, out-of-distribution drift, and gradient pathologies. Guided by findings in the original CQL~\cite{cql} paper and d3rlpy~\cite{d3rlpy} documentation, we ran 100 Optuna trials over standard ranges and selected the configuration in Table~\ref{tunetable}.

To prevent degradation or policy collapse, we employ three broad categories of early stopping checks evaluated every epoch, plus a simple sanity check on value magnitudes. If early stopping does not trigger, training proceeds to 20 epochs, based on prior trials indicating convergence by that point, to produce a full model for routing validation discussed below.

\noindent\textbf{Loss explosion or divergence detection.}
We monitor critic loss, actor loss, conservative loss, and TD error for sustained growth that signals failure in the Q approximation or unstable bootstrapping. We allow transient overshoot early in training but stop when any tracked loss exceeds $100\times$ its initial value or shows a persistent upward trend across 10 consecutive epochs. This prevents unreliable actor gradients once the critic destabilizes.

\noindent\textbf{Action difference monitoring.}
We track the d3rlpy \texttt{action\_diff} metric,
with actions normalized to $[0,1]$ during training. Moderate deviation from dataset actions is expected, but large divergence indicates poor generalization. Values above $1.0$ act as a warning threshold; we stop when the metric continues to increase across successive checkpoints.



\begin{figure}
    \centering
    \includegraphics[width=0.8\linewidth]{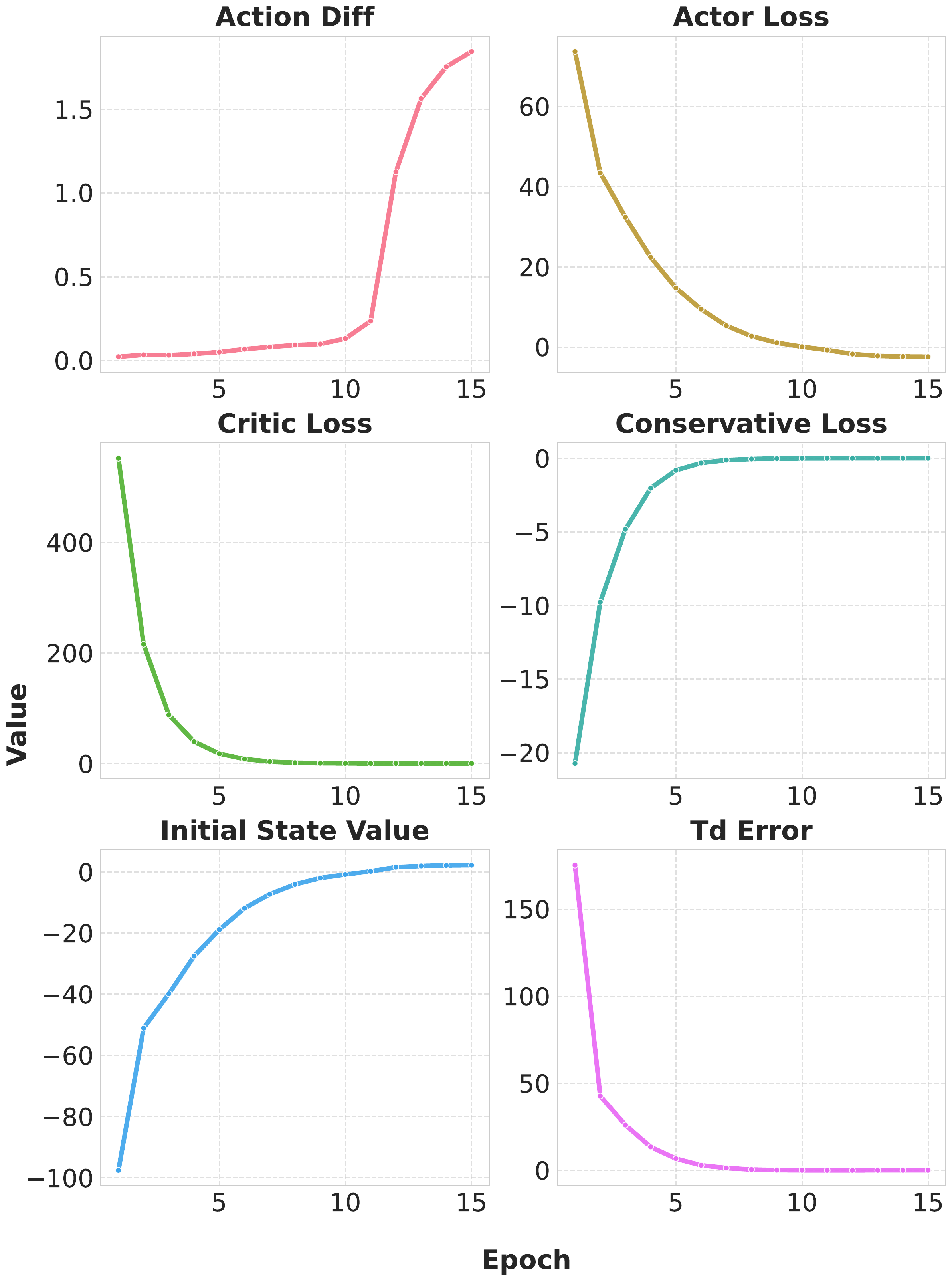}
    \caption{Model Training Progress. 1 epoch = 10,000 timesteps} \vspace{-0mm}
    \label{fig:trainingProgress}
\end{figure}

Figure~\ref{fig:trainingProgress} shows training metrics for the LSTM–CQL policy. Actor and critic losses drop sharply and then flatten near zero without overshoot. TD error decays rapidly and remains low. The conservative loss rises monotonically from a large negative value towards zero as the policy places more mass on actions supported by the dataset. The initial state value increases smoothly and plateaus; there is no overshoot. 



Here, an \emph{epoch} is a fixed budget of \emph{10,000} policy/critic update steps sampled (with replacement) from the replay buffer, not a full pass over the dataset. 
All training was performed on CPU (320 cores @ 2.10 GHz) as per-epoch runtime was comparable to an H100, likely because the high core count removed the compute bottleneck for our data and sequence models.

\noindent\textbf{Inference and integration.}
At runtime, the router calls the policy once per iteration boundary, provides the current state \(s_k\), and applies the returned multipliers in iteration \(k{+}1\). Integration touches only these 4 knobs. The underlying search remains unchanged, which keeps the approach portable across designs, densities, and adjustment settings. Inference is integrated via TorchScript/LibTorch inside the C++ router.

\section{Experimental Results}
\label{sec:ExperimentalResults}
\subsection{Methodology}
\begin{figure}[htbp]
    \centering
    
    \begin{subfigure}{0.4\columnwidth}
        \includegraphics[width=\linewidth]{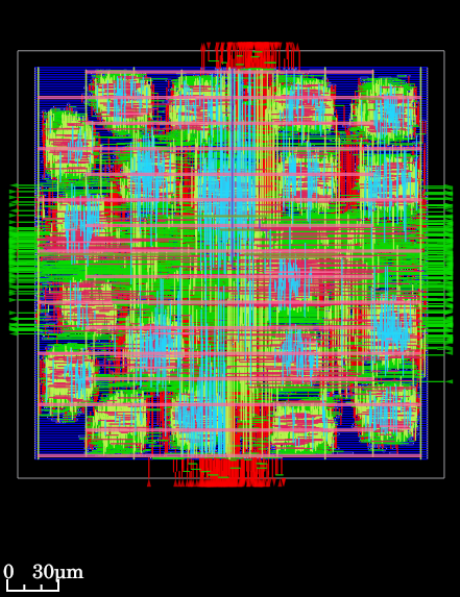}
        \caption{Density=0.42}
    \end{subfigure}%
    \hspace{0.5cm}
    \begin{subfigure}{0.3975\columnwidth}
        \includegraphics[width=\linewidth]{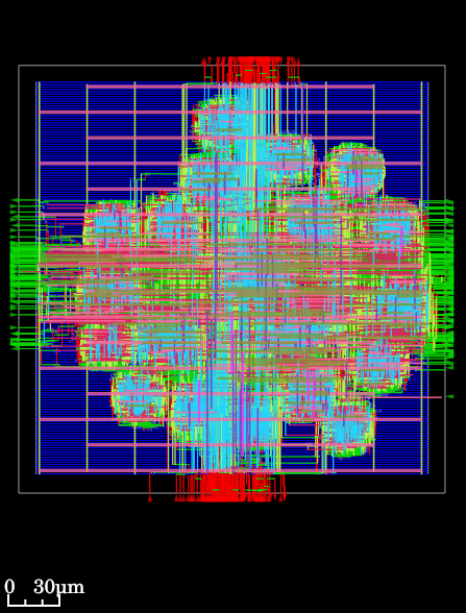}
        \caption{Density=0.65}
    \end{subfigure}\vspace{-0mm}
    \caption{Example layout (\texttt{aes}) with varying density. Core area is the same in (a) and (b), but clustering is tighter in (b).}\vspace{-0mm}
    \label{fig:5}
\end{figure}

We evaluate on OpenROAD design-suite circuits in the Nangate45 node so that placement utilization and routing adjustment can be varied. ISPD `18 and ISPD `19 benchmarks ship fixed LEF/DEF and do not permit density sweeps, so they are not used for the high utilization study here. We consider two test regimes:

\noindent\textbf{(A) Converging cases.}
The baseline converges to zero DRV within the standard cap, and we test whether the policy converges faster and in fewer iterations across \emph{unseen} operating points. To induce a distribution shift, densities at test time are interleaved between the training densities (e.g., midpoints rather than the exact values used for training), while routing adjustment is fixed at a moderately difficult setting. The density sweeps in Figure~\ref{fig:densitysweep} therefore form the core evaluation for these converging cases. By holding adjustment fixed, we isolate placement density as the primary variable and can measure how routing behavior and policy performance change with it. We also include a visual comparison of the \emph{same} region of a design at two densities in Figure~\ref{fig:5} to show that even moderate density changes can meaningfully alter topology, clustering, and blockage patterns.

\noindent\textbf{(B) Hard cases (baseline non-convergent).}
The baseline fails to reach zero DRV within 65 iterations. We report whether the policy (i) converges fully, or (ii) reduces terminal DRV versus the baseline when convergence is not reached under the same cap. These are deliberately harder operating points than those seen in training due to runtime constraints in data generation. In practice, many such failures occur at the lowest adjustment (e.g., \(0.0\)) for higher densities; some designs exhibit failure at higher adjustments as well.

\subsection{Reporting format}
For Case~(A), we present a density sweep as a plot of iterations versus placement density with two curves (baseline and RL agent). Runtime is summarized separately using a paired scatter of baseline versus policy per density on the same axis. Following the reporting used in prior work~\cite{khan2026accelerating}, Table~\ref{tab:default_density} reports metrics for a single density per design to enable direct comparison between our model and prior work. 

For Case~(B), we present Table~\ref{tab:augmented_density} with the terminal DRV count at the iteration cap; zero DRVs mean that convergence was achieved within the iteration cap. We also report runtime and wirelength. All inference overheads are included in policy runtimes. 

Our benchmarks were run on an AMD EPYC 9275F CPU @ 4.1 GHz with 768 GB of DDR5 RAM and 48 threads.
The runtime for each benchmark was averaged over 10 runs for each configuration (default and RL-guided).
Runtimes for our approach include policy inference overhead, which was measured to be about 3s total per run due to the LSTM overhead.

\subsection{Benchmark Performance}


\begin{figure}
    \centering
\includegraphics[width=1\linewidth]{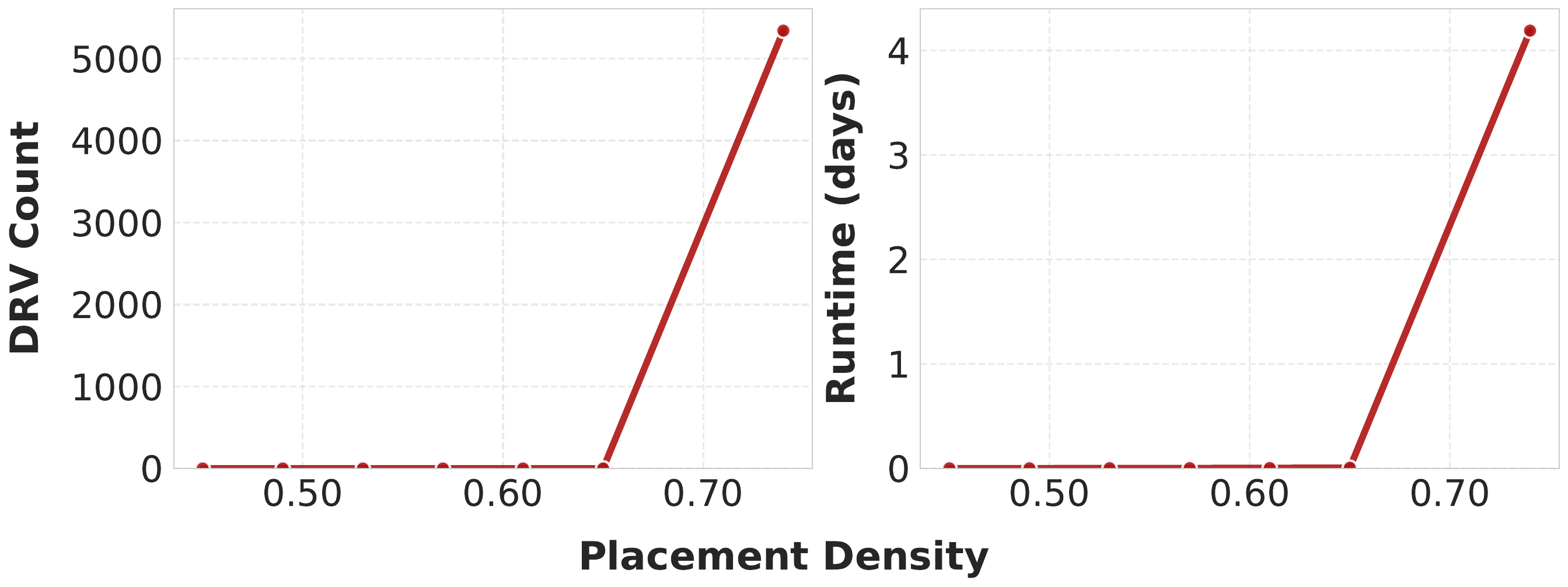}
    \caption{DRV and Runtime in \texttt{aes}} \vspace{-0mm}
    \label{fig:explosion}
\end{figure}

\begin{figure*}
    \centering
    \includegraphics[width=1\linewidth]{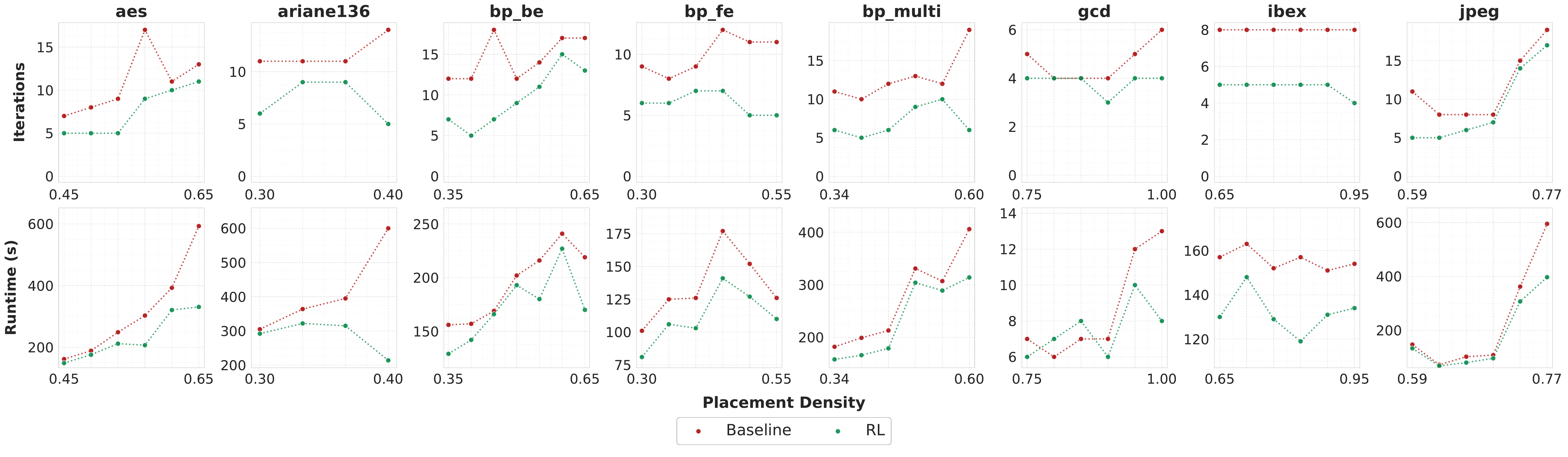}
    \caption{Density vs. routing iterations and runtime on dense benchmark designs. All designs finish with 0 DRVs.}\vspace{-0mm}
    \label{fig:densitysweep}\vspace{-0mm}
\end{figure*}

\begin{table*}
 \begin{center}  
\caption{Performance on OpenROAD Design Suite (default density. Values from the prior work~\cite{khan2026accelerating} are taken from their published results; our model is evaluated on the same test set.)\vspace{-0mm}}
\label{tab:default_density}
    \begin{tabular}{|r||r||r|r|r||r|r|r|r||r|r|r|r|} \hline 
          \textbf{Design} & \multicolumn{1}{|c||}{\textbf{Density}} &  \multicolumn{3}{|c||}{\textbf{Iterations}} & \multicolumn{4}{|c||}{\textbf{Runtime (s)}}  &\multicolumn{4}{|c|}{\textbf{Wirelength (um)}}\\ \hline 
    & & \textbf{Base} & \textbf{\cite{khan2026accelerating}}& \textbf{Ours}& \textbf{Base} & \textbf{\cite{khan2026accelerating}}&  \textbf{Ours}&\textbf{Diff}
           & \textbf{Base} & \textbf{\cite{khan2026accelerating}}& \textbf{Ours}&\textbf{Diff} \\ \hline 
 \texttt{aes}       & 0.53 &  6 &  5 &  \textbf{4} &  55 &  46 &  \textbf{42} &  -8.70\% &   309750 &   312744 &   \textbf{304773} & -2.55\% \\ \hline 
 \texttt{ariane136} & 0.30 &  6 &  \textbf{4} &  \textbf{4} & 147 & 143 & \textbf{130} &  -9.09\% &  \textbf{8017226} &  8038308 &  8038338 &  0.00\% \\ \hline 
 \texttt{bp\_be}    & 0.36 & 15 & 11 &  \textbf{9} & 115 &  56 &  \textbf{48} & -14.29\% &  3028121 &  3037392 &  \textbf{3014477} & -0.75\% \\ \hline 
 \texttt{bp\_fe}    & 0.31 & 13 &  7 &  \textbf{4} &  80 &  43 &  \textbf{36} & -16.28\% &  2352172 &  2359893 &  \textbf{2266409} & -3.96\% \\ \hline 
 \texttt{bp\_multi} & 0.35 & 11 &  \textbf{5} &  \textbf{5} & 132 & 110 & \textbf{109} &  -0.91\% &  \textbf{4711286} &  4729301 &  4729973 &  0.01\% \\ \hline 
 \texttt{gcd}       & 0.66 &  4 &  \textbf{3} &  \textbf{3} &   \textbf{3} &   4 &   \textbf{3} & -25.00\% & \textbf{4786} &     4827 &     4791 & -0.75\% \\ \hline 
 \texttt{ibex}      & 0.61 &  5 &  \textbf{4} &  \textbf{4} &  27 &  28 &  \textbf{22} & -21.43\% & \textbf{330625} &   332567 &   332558 &  0.00\% \\ \hline 
 \texttt{jpeg}      & 0.57 &  5 &  \textbf{4} &  \textbf{4} &  39 &  37 &  \textbf{30} & -18.92\% &  \textbf{1163854} &  1169491 &  1164986 & -0.39\% \\ \hline\hline
 \textbf{Total}     &  --  & 65 &        43   & \textbf{37} & 598 & 467 & \textbf{420} & -10.06\% & 19917820 & 19984523 & \textbf{19856305} & -0.64\% \\ \hline
 \end{tabular}
 \vspace{-0mm}
\end{center}
\end{table*}

\begin{table*}
 \begin{center}
\caption{Performance on OpenROAD Design Suite (augmented density). Prior work~\cite{khan2026accelerating} omitted; see Section 5.3. \vspace{-0mm}}
\label{tab:augmented_density}
    \begin{tabular}{|r||r||r|r|r||r|r|r||r|r|r|} \hline 
          \textbf{Design} & \multicolumn{1}{c||}{\textbf{Density}} &  \multicolumn{3}{|c||}{\textbf{DRVs}} & \multicolumn{3}{c||}{\textbf{Runtime (s)}}& \multicolumn{3}{c|}{\textbf{Wirelength (um)}}\\ \hline 
    && \textbf{Base} & \textbf{Ours} &\textbf{Diff}& \textbf{Base} & \textbf{Ours}&\textbf{Diff}& \textbf{Base} &\textbf{Ours}&\textbf{Diff}\\ \hline 
 \texttt{aes}        & 0.74 &   5342 &  \bf{486} & -90.90\% &  362000 & \textbf{349040} &-3.58\%& \textbf{371455} & 375030&0.96\%\\ \hline 
  \texttt{ariane136} & 0.52 & \bf{0} &    \bf{0} &       -- &     464 &    \textbf{232} & -50.00\% & \textbf{6628779} & 6669617 & 0.62\%\\ \hline
 \texttt{bp\_be}     & 0.70 &   1856 &   \bf{37} & -98.01\% &  869618 & \textbf{800179} &-7.99\%& 2698034&\textbf{2627586}&-2.61\%\\ \hline 
 \texttt{bp\_fe}     & 0.62 &   6882 &  \bf{542} & -92.12\% &  402020 & \textbf{285869} &-28.89\%& \textbf{1871959}&1905874&1.81\%\\ \hline 
 \texttt{bp\_multi}  & 0.52 &   1086 &   \bf{89} & -91.80\% &  399180 & \textbf{379243} &-4.99\%& \textbf{3848245}&3890018&1.09\%\\ \hline 
 \texttt{gcd}        & 1.00 & \bf{0} &    \bf{0} &       -- &      10 &     \textbf{11} & 10.00\%& \textbf{5326}& 5372&0.86\%\\\hline \hline
 \texttt{ibex}       & 1.00 & \bf{0} &    \bf{0} &       -- &     166 &    \textbf{146} & -12.05\%& 333569& \textbf{330897}&-0.80\%\\ \hline
 \texttt{jpeg}       & 0.80 &    601 &   \bf{61} & -89.85\% &  126000 & \textbf{119700} &-5.00\%& \textbf{1046000}&1052050&0.58\%\\ \hline

 \textbf{Total}      &  --  &  15767 & \bf{1215} & -92.29\% & 2159458 &  \textbf{1934420} & -10.42\% & \textbf{16803367} &  16856444 & 0.32\%\\ \hline
 \end{tabular}
 \vspace{-0mm}
\end{center}
\end{table*}



Our experimental results are summarized in Tables \ref{tab:default_density} and \ref{tab:augmented_density}. We begin with the converging cases. Table~\ref{tab:default_density} reports a single density per design across 1) the baseline OpenROAD router 2) reported values from prior work~\cite{khan2026accelerating} 3) our model. All three are compared on identical designs at the same operating point. Figure~\ref{fig:densitysweep} then extends this to full density sweeps. For most designs, the iteration and runtime improvements from our model persist across held-out densities. While each design in this set is included in the training data, the specific density and adjustment configurations tested here were \emph{not} present during training. Visual inspection of the generated layouts (Figure~\ref{fig:5}) confirms differences in placement topology between densities. Therefore, the results demonstrate the model's ability to generalize across unseen routing settings which can induce dramatically different layout topologies on the same design rather than simply memorizing training examples. 

We then present Table~\ref{tab:augmented_density} to show results across 1) the baseline OpenROAD detailed router and 2) our RL model for the harder non-converging cases. We omit  prior work \cite{khan2026accelerating} here because it reports a single placement density and does not perform sweeps. Sweeping density or adjustment drastically shifts the distribution of the raw features a learned policy consumes, and a policy trained at one density on those raw features saturates under the shift. This out-of-distribution behavior is well documented by several works. Dakhmouche and Gorji~\cite{dakhmouche2025neuralnetworksmasterextrapolation} examine why machine learning models fail to extrapolate beyond their training distribution. Figure~\ref{fig:model_comparison} further corroborates our claim that the prior approach~\cite{khan2026accelerating} degrades sharply as density rises while our model with recurrent memory remains stable. The baseline, by contrast, is an expert curated cost schedule, not a learned policy, so it does not suffer such saturation. Using machine learning theory and Figure~\ref{fig:model_comparison}, we can accurately predict where the prior approach~\cite{khan2026accelerating} degrades, but the same cannot be approximated for a human written baseline. Thus, we compare our performance with the static baseline that the prior approach~\cite{khan2026accelerating} itself builds on.

Table~\ref{tab:augmented_density} also reports substantially higher runtimes than Table~\ref{tab:default_density} since these cases route under minimal guide quality (adjustment=0.0), where the detailed router has fewer reserved resources and must rip-up and reroute extensively, as discussed in Section~\ref{sec:Methodology}. Figure~\ref{fig:explosion} shows this runtime explosion when entering the hard regime in \texttt{aes}. The only exceptions we found were \texttt{ibex} and \texttt{gcd}, which maintained reasonable runtime under extreme conditions due to their simpler routing topologies.

Across both regimes, our model consistently reduces iterations and runtime in the converging cases and significantly reduces DRVs in the non-converging cases. We see a 92\% decrease in DRVs from the baseline alongside a 10\% runtime improvement. Three cases in Table \ref{tab:augmented_density} stand out. \texttt{ibex} and \texttt{gcd} were evaluated at the maximum possible density with the hardest adjustment settings but still converged. Notably, our model still achieves improvements over the baseline. For \texttt{ariane136}, we were unable to identify a non-converging operating point within reasonable runtime constraints and therefore report results at the highest utilization settings we explored.

For \texttt{gcd} and \texttt{ibex}, the RL inference overhead dominates over the speedup from RL and thus total changes remain modest. However, in harder cases (e.g., higher cell count or extreme routing congestion) that motivate this work, the inference cost is well under 0.01\% of total routing time and is negligible against the time this policy saves.

Overall wirelength impact remains small, with a 0.32\% total increase across the augmented density set. Any increases in wirelength over 1\% are balanced by reducing DRV count over the baseline by \textgreater{}90\%. We did not include wirelength in the reward function and therefore it was not an optimization target when selecting weights.

\section{Conclusion}
\label{sec:Conclusion}

This paper presented a history aware RL agent for iterative cost control under varying routing conditions for improved detailed routing convergence and quality. We introduce a novel offline RL model with LSTM to implement our methodology that can be extended to any router with iterative search.
Our model integrates design features and DRV history to infer the router's cost weights, making it portable and adaptable to any router which uses iterative cost-based algorithms. We evaluated 8 designs with augmented densities from the OpenROAD Design Suite, demonstrating an average 92\% DRV reduction while achieving an average 10\% runtime improvement over the baseline. Compared to the prior CQL-only approach \cite{khan2026accelerating}, our LSTM-enhanced architecture better captures routing dynamics and improves convergence across densities.

\section*{Acknowledgments}
We thank Saik Anam Siam for helpful guidance on model saturation from distribution shift in machine learning. This work was partially funded by a Google charitable gift.

\bibliographystyle{IEEEtran}
\bibliography{software}

@string{iccad = "Proceedings of the IEEE/ACM International Conference on
	Computer-Aided Design"}

@string{dac = "Proceedings of the ACM/IEEE Design Automation Conference"}

@string{aspdac = "Proceedings of the Asia-South Pacific Design Automation
	Conference"}

@string{dateconf = "Proceedings of Design, Automation \& Test in Europe"}

@string{ispd = "Proceedings of the International Symposium on Physical Design"}

@string{iccad = "Proceedings of the IEEE/ACM International Conference on Computer-Aided Design"}

@string{aspdac = "Proceedings of the Asia-South Pacific Design Automation Conference"}

@string{dateconf = "Proceedings of the Design, Automation \& Test in Europe"}

@string{nips="Advances in Neural Information Processing Systems"}

@string{iccad = " Proc. ICCAD"}

@string{dac = " Proc. DAC"}

@string{aspdac = " Proc. ASP-DAC"}

@string{dateconf = " Proc. DATE "}

@string{ispd = " Proc. ISPD"}

@string{nips = "Proc. NeurIPS"}

@inproceedings{ajayi2019openroad-gomactech,
  title={{OpenROAD: Toward a Self-Driving, Open-Source Digital Layout Implementation Tool Chain}},
  author={Ajayi, Tutu and Blaauw, David and Chan, Tuck-Boon and Cheng, Chung-Kuan and Chhabria, Vidya A. and Choo, David K. and Coltella, Matteo and Dobre, Sorin and Dreslinski, Ronald G. and Fogaça, Mateus and Hashemi, Soheil and Hosny, Abdelrahman and Kahng, Andrew B. and Kim, Minsoo and Li, Jiajia and Liang, Zhaoxin and Mallappa, Uday and Penzes, Paul and Pradipta, Geraldo and Reda, Sherief and Rovinski, Austin and Samadi, Kambiz and Sapatnekar, Sachin S. and Saul, Lawrence and Sechen, Carl and Srinivas, Vaishnav and Swartz, William and Sylvester, Dennis and Urquhart, David and Wang, Lutong and Woo, Mingyu and Xu, Bangqi},
  booktitle = {Proceedings of Government Microcircuit Applications and Critical Technology Conference},
  series = {GOMACTech '19},
  year = {2019}
}

@inproceedings{rovinski2020bridging,
 author = {Rovinski, Austin and Ajayi, Tutu and Kim, Minsoo and Wang, Guanru and Saligane, Mehdi},
 title = {{Bridging Academic Open-Source EDA to Real-World Usability}},
 booktitle = iccad,
 pages = {1--7},
 year = {2020}
}

@article{kahng2020tritonroute,
  title={{TritonRoute: The Open-Source Detailed Router}},
  author={Kahng, Andrew B and Wang, Lutong and Xu, Bangqi},
  journal={IEEE Transactions on Computer-Aided Design of Integrated Circuits and Systems},
  volume={40},
  number={3},
  pages={547--559},
  year={2020},
  publisher={IEEE}
}

@inproceedings{kahng2018tritonroute,
  title={{TritonRoute: An Initial Detailed Router for Advanced VLSI Technologies}},
  author={Kahng, Andrew B and Wang, Lutong and Xu, Bangqi},
  booktitle = iccad,
  pages={1--8},
  year={2018},
  organization={IEEE}
}

@ARTICLE{lee1961algorithm,
  author={Lee, C. Y.},
  journal={IRE Transactions on Electronic Computers}, 
  title={{An Algorithm for Path Connections and Its Applications}}, 
  year={1961},
  volume={EC-10},
  number={3},
  pages={346-365},
  doi={10.1109/TEC.1961.5219222}}

@inproceedings{cql,
 author = {Kumar, Aviral and Zhou, Aurick and Tucker, George and Levine, Sergey},
 booktitle = nips,
 pages = {1179--1191},
 publisher = {Curran Associates, Inc.},
 title = {{Conservative Q-Learning for Offline Reinforcement Learning}},
 volume = {33},
 year = {2020}
}

@INPROCEEDINGS{10,
  author={Zeng, Wei and Davoodi, Azadeh and Topaloglu, Rasit Onur},
  booktitle=dateconf, 
  title={{Explainable DRC Hotspot Prediction with Random Forest and SHAP Tree Explainer}}, 
  year={2020},
  volume={},
  number={},
  pages={1151-1156},
  doi={10.23919/DATE48585.2020.9116488}}

@ARTICLE{11,
  author={Park, Hyunbum and Baek, Kyeonghyeon and Kim, Suwan and Choi, Kyumyung and Kim, Taewhan},
  journal={IEEE Transactions on Computer-Aided Design of Integrated Circuits and Systems}, 
  title={{Pin Accessibility and Routing Congestion Aware DRC Hotspot Prediction for Designs in Advanced Technology Nodes with Consolidated Practical Applicability and Sustainability}}, 
  year={2024},
  volume={43},
  number={12},
  pages={4786-4799},
  doi={10.1109/TCAD.2024.3405894}}

@inproceedings{12,
    author = {Liang, Rongjian and Xiang, Hua and Pandey, Diwesh and Reddy, Lakshmi and Ramji, Shyam and Nam, Gi-Joon and Hu, Jiang},
    title = {{DRC Hotspot Prediction at Sub-10nm Process Nodes Using Customized Convolutional Network}},
    year = {2020},
    isbn = {9781450370912},
    publisher = {Association for Computing Machinery},
    address = {New York, NY, USA},
    doi = {10.1145/3372780.3375560},
    booktitle = ispd,
    pages = {135–142}
}

@inproceedings{13,
author = {Chen, Hao and Hsu, Kai-Chieh and Turner, Walker J. and Wei, Po-Hsuan and Zhu, Keren and Pan, David Z. and Ren, Haoxing},
title = {{Reinforcement Learning Guided Detailed Routing for Custom Circuits}},
year = {2023},
isbn = {9781450399784},
publisher = {Association for Computing Machinery},
address = {New York, NY, USA},
doi = {10.1145/3569052.3571874},
booktitle = ispd,
pages = {26–34},
numpages = {9},
location = {Virtual Event, USA},
series = {ISPD '23}
}

@article{bidir,
author = {Kaindl, Hermann and Kainz, Gerhard},
title = {{Bidirectional Heuristic Search Reconsidered}},
year = {1997},
issue_date = {July 1997},
publisher = {AI Access Foundation},
address = {El Segundo, CA, USA},
volume = {7},
number = {1},
issn = {1076-9757},
journal = {J. Artif. Int. Res.},
month = dec,
pages = {283–317},
numpages = {35}
}

@inproceedings{arnold,
author = {Arnold, Michael H. and Scott, Walter S.},
title = {{An Interactive Maze Router with Hints}},
year = {1988},
isbn = {0818688645},
publisher = {IEEE Computer Society Press},
address = {Washington, DC, USA},
booktitle = dac,
pages = {672–676},
numpages = {5},
location = {Atlantic City, New Jersey, USA},
series = {DAC '88}
}

@inproceedings{han2015,
author = {Han, Kwangsoo and Kahng, Andrew B. and Lee, Hyein},
title = {{Evaluation of BEOL Design Rule Impacts Using an Optimal ILP-Based Detailed Router}},
year = {2015},
isbn = {9781450335201},
publisher = {Association for Computing Machinery},
address = {New York, NY, USA},
doi = {10.1145/2744769.2744839},
booktitle = dac,
articleno = {68},
numpages = {6},
location = {San Francisco, California},
series = {DAC '15}
}

@inproceedings{nieberg2011,
author = {Nieberg, Tim},
title = {{Gridless Pin Access in Detailed Routing}},
year = {2011},
isbn = {9781450306362},
publisher = {Association for Computing Machinery},
address = {New York, NY, USA},
doi = {10.1145/2024724.2024763},
booktitle = dac,
pages = {170–175},
numpages = {6},
location = {San Diego, California},
series = {DAC '11}
}

@ARTICLE{ding2017,
  author={Ding, Yixiao and Chu, Chris and Mak, Wai-Kei},
  journal={IEEE Transactions on Computer-Aided Design of Integrated Circuits and Systems}, 
  title={{Self-Aligned Double Patterning Lithography Aware Detailed Routing with Color Preassignment}}, 
  year={2017},
  volume={36},
  number={8},
  pages={1381-1394},
  doi={10.1109/TCAD.2016.2622625}
}

@INPROCEEDINGS{liu2014,
  author={Iou-Jen Liu and Shao-Yun Fang and Yao-Wen Chang},
  booktitle = dac, 
  title={{Overlay-Aware Detailed Routing for Self-Aligned Double Patterning Lithography Using the Cut Process}}, 
  year={2014},
  volume={},
  number={},
  pages={1-6},
  doi={10.1109/DAC.2014.6881377}}

@article{ahrens2015detailed,
author = {Ahrens, Markus and Gester, Michael and Klewinghaus, Niko and Muller, Dirk and Peyer, Sven and Schulte, Christian and Tellez, Gustavo},
title = {{Detailed Routing Algorithms for Advanced Technology Nodes}},
year = {2015},
issue_date = {April 2015},
publisher = {IEEE Press},
volume = {34},
number = {4},
issn = {0278-0070},
doi = {10.1109/TCAD.2014.2385755},
journal = {IEEE Transactions on Computer-Aided Design of Integrated Circuits and Systems},
month = apr,
pages = {563--576},
numpages = {14}
}

@inproceedings{hightower1969solution,
author = {Hightower, David W.},
title = {{A Solution to Line-Routing Problems on the Continuous Plane}},
year = {1969},
isbn = {9781450379298},
publisher = {Association for Computing Machinery},
address = {New York, NY, USA},
doi = {10.1145/800260.809014},
booktitle = dac,
pages = {1–24},
numpages = {24},
series = {DAC '69}
}

@article{d3rlpy,
  author = {Seno, Takuma and Imai, Michita},
  title = {{d3rlpy: An Offline Deep Reinforcement Learning Library}},
  journal = {Journal of Machine Learning Research},
  volume = {23},
  year = {2022}
}

@inproceedings{chen2020detailed,
  author = {Chen, Gengjie and Pui, Chak-Wa and Li, Haocheng and Chen, Jingsong and Jiang, Bentian and Young, Evangeline F. Y.},
  title = {{Detailed Routing by Sparse Grid Graph and Minimum-Area-Captured Path Search}},
  year = {2019},
  isbn = {9781450360074},
  publisher = {Association for Computing Machinery},
  address = {New York, NY, USA},
  doi = {10.1145/3287624.3287678},
  booktitle = aspdac,
  pages = {754–760},
  numpages = {7},
  location = {Tokyo, Japan},
  series = {ASPDAC '19}
}

@inproceedings{mantik2018ispd,
author = {Mantik, Stefanus and Posser, Gracieli and Chow, Wing-Kai and Ding, Yixiao and Liu, Wen-Hao},
title = {{ISPD 2018 Initial Detailed Routing Contest and Benchmarks}},
year = {2018},
isbn = {9781450356268},
publisher = {Association for Computing Machinery},
address = {New York, NY, USA},
doi = {10.1145/3177540.3177562},
booktitle = ispd,
pages = {140–143},
numpages = {4},
location = {Monterey, California, USA},
series = {ISPD '18}
}

@inproceedings{liu2019ispd,
author = {Liu, Wen-Hao and Mantik, Stefanus and Chow, Wing-Kai and Ding, Yixiao and Farshidi, Amin and Posser, Gracieli},
title = {{ISPD 2019 Initial Detailed Routing Contest and Benchmark with Advanced Routing Rules}},
year = {2019},
isbn = {9781450362535},
publisher = {Association for Computing Machinery},
address = {New York, NY, USA},
doi = {10.1145/3299902.3311067},
booktitle = ispd,
pages = {147–151},
numpages = {5},
location = {San Francisco, CA, USA},
series = {ISPD '19}
}

@article{featuremask,
  title={{CITADEL: A Semi-Supervised Active Learning Framework for Malware Detection under Continuous Distribution Drift}}, 
  author={Md Ahsanul Haque and Md Mahmuduzzaman Kamol and Ismail Hossain and Suresh Kumar Amalapuram and Vladik Kreinovich and Mohammad Saidur Rahman},
  year={2025},
  journal={arXiv preprint \url{https://arxiv.org/abs/2511.11979}}
}

@INPROCEEDINGS{khan2026accelerating,
  author={Khan, Afsara and Rovinski, Austin},
  booktitle=dateconf, 
  title={{Accelerating Detailed Routing Convergence through Offline Reinforcement Learning}}, 
  year={2026},
  volume={},
  number={},
  pages={1-7},
  doi={10.23919/DATE69613.2026.11539594}}

@article{dakhmouche2025neuralnetworksmasterextrapolation,
  title={{Why Cannot Neural Networks Master Extrapolation? Insights from Physical Laws}}, 
  author={Ramzi Dakhmouche and Hossein Gorji},
  year={2025},
  journal={arXiv preprint \url{https://arxiv.org/abs/2510.04102}}
}



\end{document}